\documentclass{sig-alternate-05-2015}

\usepackage{hyperref}
\usepackage{graphicx}
\usepackage{amsmath}
\newtheorem{defn}{Definition}[section]
\newtheorem{exmp}{Example}[section]
\newtheorem{prop}{Proposition}[section]
\usepackage{amssymb}
\usepackage{subfigure}
\usepackage{ifsym}
\usepackage{stmaryrd}
\usepackage{paralist}
\usepackage{algorithm}
\usepackage{algorithmic}
\usepackage{balance}
\usepackage{tikz}
\usepackage{url}
\usepackage{verbatim}
\usetikzlibrary{arrows,positioning,automata,decorations,fit,backgrounds,calc,shapes,snakes}
\usetikzlibrary{shapes.geometric} 
\usetikzlibrary{decorations.pathmorphing} 
\usepackage{xcolor}
\usepackage{pgfplots}
\usetikzlibrary{matrix}
\newcommand{\ANDAND}{\mathbin{\mathrm{AND}}}
\newcommand{\OPT}{\mathbin{\mathrm{OPT}}}

\newcommand{\UNION}{\mathbin{\mathrm{UNION}}}

\newcommand{\FILTER}{\mathbin{\mathrm{FILTER}}}
\newcommand{\SELECT}{\mathop{\mathrm{SELECT}}}

\newcommand{\kst}{\mathit{KST}}

\newcommand{\var}{\mathit{var}}
\newcommand{\ld}{\mathit{LD}}

\newcommand{\true}{\mathit{true}}
\newcommand{\false}{\mathit{false}}
\newcommand{\error}{\mathit{error}}

\newcommand{\semm}[2]{\llbracket #1 \rrbracket_{#2}}

\newcommand{\dom}[1]{\mathrm{dom}(#1)}

\newcommand{\BGP}[1]{\mathit{BGP}(#1)}
\newcommand{\lev}[1]{\mathit{level}(#1)}

\newcommand{\LT}[1]{\mathit{LT}(#1)}
\newcommand{\LM}[1]{\mathit{LM}(#1)}

\newcommand{\depth}[1]{\mathit{dep}(#1)}
\def\sharedaffiliation{%
\end{tabular}
\begin{tabular}{c}}

\begin{document}

\setcopyright{acmcopyright}

\doi{xxx.xxx/xxx}

\isbn{xxx-xxxx-xx-xxxx}



%

\title{Efficient Approximation of Well-Designed SPARQL Queries}
\numberofauthors{5}
\author{
      \alignauthor Zhenyu Song\\
      \email{szyhw@tju.edu.cn}
      \alignauthor Zhiyong Feng\\
      \email{zyfeng@tju.edu.cn}
      \alignauthor Xiaowang Zhang\\
      \email{xiaowangzhang@tju.edu.cn}
\and
      \alignauthor Xin Wang\\
      \email{wangx@tju.edu.cn}
      \alignauthor Guozheng Rao\\
      \email{rgz@tju.edu.cn}
      \sharedaffiliation
      \affaddr{School of Computer Science and Technology, Tianjin University, Tianjin, China}  \\
      \affaddr{Tianjin Key Laboratory of Cognitive Computing and Application, Tianjin, China}
}
%

\maketitle
\begin{abstract}
Query response time often influences user experience in the real world.
However, it possibly takes more time to answer a query with its all exact solutions, especially when it contains the OPT operations since
the OPT operation is the least conventional operator in SPARQL. So it
becomes essential to make a trade-off between the query response time and the accuracy of their solutions. In this paper, based on the
depth of the OPT operation occurring in a query, we propose an approach
to obtain its all approximate queries with less depth of the OPT
operation. This paper mainly discusses those queries with well-designed
patterns since the OPT operation in a well-designed pattern is really
``optional''. Firstly, we transform a well-designed pattern in OPT normal form into a
well-designed tree, whose inner nodes are labeled by OPT operation and
leaf nodes are labeled by patterns containing other operations such as the AND operation and the FILTER
operation. Secondly, based on this well-designed tree, we remove
``optional'' well-designed subtrees with less depth of the OPT
operation and then obtain approximate queries with different depths of the
OPT operation. Finally, we evaluate the approximate query efficiency with the degree of
approximation.
\end{abstract}

%
%


%
%

%
%



\section{Introduction}\label{sec:introduction}
Currently, there is renewed interest in the classical topic of graph
databases \cite{gutierrez_survey,wood_survey,walklogic}. Much of this
interest has been sparked by SPARQL: the query language for RDF\@. Resource
Description Framework (RDF) \cite{Klyne2004Resource} is the standard data
model in the Semantic Web. RDF describes the relationship of entities or
resources using directed label graph. RDF has a broad range of applications
in the Semantic Web, social network, bio-informatics, geographical data, etc
\cite{abiteboul2000data}. An example in Table \ref{professor.rdf} has been
given to describe the entities of \textit{Jon Smith} and \textit{Liz Ben}. For
example, in the first line, it describes that the person \textit{Jon smith}
works for \textit{Semantic University}. {SPARQL} \cite{Prud2007SPARQL}
recommended by W3C has become the standard language for querying RDF data
since 2008 by inheriting classical relational languages such as SQL.
\vspace{-3mm}
\begin{table}[h]
\centering
\caption{professor.rdf \label{professor.rdf}}
\begin{tabular}{|l|l|l|}
\hline
Jon Smith &workFor &Semantic University \\ \hline
Jon Smith &teachOf &Liz Ben \\ \hline
Jon Smith &rdf:type &professor \\ \hline
Liz Ben &rdf:type &master \\ \hline
Liz Ben &advisor &Jon Smith \\ \hline
Liz Ben &takesCourse & Ontology \\ \hline
\end{tabular}
\end{table}

In the process of information retrieval, users' tolerable waiting time is
limited \cite{Nah2003A}. Users also might have tolerable waiting time for
querying RDF data. For a SPARQL query, if it contains the OPT operation, it
will take much time to query optional pattern in SPARQL since OPT is the
least conventional operator in AND, OPT, FILTER, SELECT and UNION
\cite{Zhang2014On}. It has been shown in
\cite{P2009Semantics,schmidt_sparqloptim} that the complexity of SPARQL
query evaluation raises from PTIME-membership for the conjunctive fragment
to PSPACE-completeness when OPT operation is considered. 
So it is important to make a trade-off between query
response time and accuracy of solutions, which is a traditional topic in
databases \cite{Abiteboul1995Foundations}. Since it is hard to obtain all exact solutions of a
SPARQL query in a fixed time, a natural idea to reduce the response time
of SPARQL query is by removing some ``optional'' parts of this query (i.e.,
occurrences of the OPT operator). Moreover, we still expect
to preserve its ``non-optional'' part of this query. For instance, consider
a pattern $Q$ as follows:
\begin{multline*}
Q=((?x, \textit{rdf:type}, \textit{professor})\ \OPT \\ \ ((?x, \textit{workFor}, ?y)\ \OPT \ (?x, \textit{teachOf}, ?z))).
\end{multline*}
Here $(?x, \textit{rdf:type}, \textit{professor})$ is a ``non-optional'' pattern in
this query while both $(?x, \textit{workFor}, ?y)$ and $(?x, \textit{teachOf}, ?z))$ are ``optional'' patterns. Based on this natural idea,
there are three possible new patterns with less OPT operators as follows:
\begin{compactitem}
\item $Q_1=(?x, \textit{rdf:type}, \textit{professor})$;
\item $Q_2=((?x, \textit{rdf:type}, \textit{professor})\ \OPT \ (?x,
    \textit{workFor}, ?y))$;
\item $Q_3=((?x, \textit{rdf:type}, \textit{professor})\ \OPT \ (?x,
    \textit{teachOf}, ?z))$.
\end{compactitem}
Clearly, we can find that $Q_1$ and $Q_2$ are ideal candidates which contain
less optional patterns with protecting ``non-optional'' patterns while $Q_3$
is not since $(?x, \textit{teachOf}, ?z)$ directly depends on
$(?x, \textit{workFor}, ?y)$.

In 2015, Barcel{\'{o}}, Pichler, and Skritek \cite{barcelo2015efficient}
proposed the notion of approximation (for short, BPS's \emph{approximation})
to characterize ``\emph{partial answer}'', that is, an answer can be
extended to a ``\emph{maximal answer}'' (i.e., exact answer) of a SPARQL
query represented in well-designed pattern trees \cite{Letelier2012Static}.
In this sense, the evaluation problems of $Q_1$ and $Q_2$ are taken as the
partial evaluation problems of $Q$. However, we investigated that the BPS's
approximation did not provide a fine-grained classification between $Q_1$
and $Q_2$. For users, they can't judge which one will lead to less query response time within tolerable waiting time.

In this paper, based on the depth of OPT operation occurring in a query,
we propose an approach to obtain its all approximate queries with less
depth of the OPT operation. We mainly
consider the fragment of UNION-free well-designed SPARQL patterns where the
OPT operator is a really ``optional'' operation \cite{P2009Semantics}.
Besides, the UNION-free well-designed SPARQL fragment is indeed
maximal among all fragments of LSQ \cite{saleem2015lsq} - a linked dataset describing SPARQL
queries extracted from the logs of public SPARQL endpoints in our real world. For simplification, we directly call
well-designed patterns instead of UNION-free well-designed SPARQL patterns.
The main contributions of this paper can be summarized as follows:
\begin{compactitem}
\item Firstly, we provide the conception of OPT-depth.
For a well-designed pattern in \emph{OPT normal form}, its OPT-depth can describe the depth of OPT operation occurring in this pattern.
Our approximation approach method is proposed based on the \emph{OPT normal form} via OPT-depth.
\item Secondly, we treat a well-designed pattern in \emph{OPT normal form} as a well-designed
    tree, whose inner nodes are labeled by OPT operation.
    We apply our approximation method by removing ``optional'' subtrees of a well-designed tree.
 \item Finally, through comparison with the non-approximate queries on LUBM dataset, the approximate queries lead to better performance.
\end{compactitem}

The rest of this paper is organized as follows: Section \ref{sec:pre}
briefly introduces the SPARQL and conception of well-designed patterns.
Section \ref{sec:approximate} defines the $k$-approximation queries. Section
\ref{sec:well-designed tree} presents the well-designed tree to capture
$k$-approximation queries and Section \ref{sec:implement} evaluates
experimental results. Finally, Section \ref{sec:con} summarizes the paper.

\section{preliminaries}\label{sec:pre}
In this section, we introduce the syntax and semantics of SPARQL 1.0 and
well-designed patterns \cite{P2009Semantics}.
\subsection{RDF} Let ${I}$, ${B}$ and ${L}$ be infinite sets of
\emph{IRIs}, \emph{blank nodes} and \emph{literals}, respectively.  These
three sets are pairwise disjoint.  We denote the union $I \cup B \cup L$ by
$U$, and elements of $I \cup L$ will be referred to as \emph{constants}.

A triple $(s, p, o) \in ({I}\cup {B}) \times {I} \times ({I} \cup
{B}\cup {L})$ is called an \emph{RDF triple}.  An \emph{RDF
graph} is a finite set of RDF triples.

\subsection{The Syntax and Semantics of SPARQL}
Assume furthermore an infinite set $V$ of \emph{variables},
disjoint from $U$. The convention is to write
variables starting with the character `?'.
SPARQL \emph{patterns} are inductively defined as follows.
\begin{compactitem}
\item Any triple from $({I}\cup {L} \cup {V}) \times ({I} \cup
{V}) \times ({I} \cup {L} \cup {V}$) is a pattern (called a
\emph{triple pattern}).
A Basic Graph Pattern (BGP) is a set of triple patterns.
\item If $P_{1}$ and $P_{2}$ are patterns, then so are
the following: $P_{1} \UNION P_{2}$, $P_{1} \ANDAND P_{2}$ and
$P_{1} \OPT P_{2}$.
\item
If $P$ is a pattern and $S$ is a finite set of variables then $\SELECT_{S}(P)$ is a pattern.
\item If $P$ is a pattern and $C$ is a constraint (defined next), then $P
    \FILTER C$ is a pattern; we call $C$ the \emph{filter condition}. Here,
    a \emph{constraint} is a boolean combination of \emph{atomic
    constraints}.
\end{compactitem}

The semantics of patterns is defined in terms of sets of
so-called \emph{mappings}, which are simply total functions
$\mu \colon S \to U$ on some finite set $S$ of variables.
We denote the domain $S$ of $\mu$ by $\dom \mu$.

Now given a graph $G$ and a pattern $P$, we define the semantics
of $P$ on $G$, denoted by $\semm P G$, as a set of mappings, in
the following manner.
\begin{compactitem}
\item If $P$ is a triple pattern $(u,v,w)$, then\\
$ \semm P G = \{\mu \colon \{u,v,w\} \cap V \to U
\mid(\mu(u),\mu(v),\mu(w)) \in G\}. $
Here, for any mapping $\mu$ and any constant $c \in I \cup L$, we
agree that $\mu(c)$ equals $c$ itself.  In other
words, mappings are extended to constants according to the
identity mapping.
\item
If $P$ is of the form $P_1 \UNION P_2$, then $\semm P G = \semm {P_1} G \cup \semm {P_2} G$.
\item
If $P$ is of the form $P_1 \ANDAND P_2$, then $\semm P G = \semm {P_1} G \Join \semm {P_2} G$,
where, for any two sets of mappings $\Omega_1$ and $\Omega_2$,
we define
$
\Omega_1 \Join \Omega = \{\mu_1 \cup \mu_2 \mid \mu_1 \in \Omega_1 \text{ and } \mu_2 \in \Omega_2 \text{ and }\mu_1 \sim \mu_2\}.
$
Here, two mappings $\mu_1$ and $\mu_2$ are called
\emph{compatible}, denoted by $\mu_1 \sim \mu_2$, if
they agree on the intersection of their domains, i.e.,
if for every variable $?x \in \dom {\mu_1} \cap \dom {\mu_2}$, we have
$\mu_1(?x) = \mu_2(?x)$.  Note that when $\mu_1$ and $\mu_2$ are
compatible, their union $\mu_1 \cup \mu_2$ is a well-defined
mapping; this property is used in the formal definition above.
\item
If $P$ is of the form $P_1 \OPT P_2$, then
$
\semm P G = (\semm {P_1} G \Join \semm {P_2} G)
\cup (\semm {P_1} G \smallsetminus \semm {P_2} G),
$
where for any two sets of mappings $\Omega_1$ and $\Omega_2$,
we define
$\Omega_1 \smallsetminus \Omega_2$ $ =
\{ \mu_1 \in \Omega_1 \mid \neg \exists \mu_2 \in \Omega_2 :
\mu_1 \sim \mu_2\}$.

\item If $P$ is of the form $\SELECT_{S}(P_{1})$, then \\ $\semm {P}{G} =
    \{\mu|_{S \cap \dom \mu} \mid \mu \in \semm{P_1}G\}$.

\item If $P$ is of the form $P_1 \FILTER C$, then $\semm P G = \{\mu \in
    \semm {P_1} G \mid \mu(C) = \true \}$. Here, for any mapping $\mu$ and constraint $C$, the evaluation of
$C$ on $\mu$, denoted by $\mu(C)$, is defined as normal in terms of a
three-valued logic with truth values $\true$, $\false$ and $\error$.
%
\end{compactitem}

\subsection{Well-designed Patterns}
The notion of well-designed patterns is introduced to characterize the
\emph{weak monotonicity} \cite{P2009Semantics}.

A $\UNION$-free pattern $P$ is well-designed if the followings hold:
\begin{compactitem}
\item $P$ is safe, that is, each subpattern of the form $Q\ \FILTER C$ of $P$
    holds the condition: $\var(C) \subseteq \var(Q)$.
\item for every subpattern $P' = (P_1 \OPT P_2)$ of $P$ and
    for every variable $?x$ occurring in $P$, the following condition hold:
If $?x$ occurs both inside $P_{2}$ and outside $P'$, then it also occurs in
$P_{1}$.
\end{compactitem}

For instance, the pattern $Q$ in Section \ref{sec:introduction} is a well-designed pattern.
However, consider the pattern $(((?x, \textit{p}, ?y) \OPT \ (?y, \textit{q}, ?z)) \OPT (?x, \textit{r}, ?z))$, it is
not a well-designed pattern since $?z$ occurs in both $(?y, \textit{q}, ?z)$ and $(?x,
\textit{r}, ?z)$ but $?z$ does not occur in $(?x, \textit{p}, ?y)$.

Note that the OPT operation provides really optional left-outer join due to
the weak monotonicity \cite{P2009Semantics}, which is an important property
to characterize the satisfiability of SPARQL \cite{Zhang2015sat}. For
instance, consider the pattern $Q$ in Section \ref{sec:introduction}, $(?x, \textit{workFor}, ?y)$
and $(?x, \textit{teachOf},?z)$ are freely optional.

\section{Approximate queries}\label{sec:approximate}
In this section, we introduce our approximation method in the \emph{OPT normal
form}.
\subsection{OPT Normal Form}
A UNION-free pattern $P$ is in \emph{OPT normal form} \cite{P2009Semantics}
if $P$ meets one of the following two conditions:
\begin{compactitem}
\item $P$ is constructed by using only the $\ANDAND$ and $\FILTER$
    operators;
\item $P = (P_1\ \OPT\ P_2)$ where $P_1$ and $P_2$ patterns are in OPT
    normal form.
\end{compactitem}

For instance, the pattern $Q$ stated in Section \ref{sec:introduction} is in OPT normal form.
However, consider the pattern $(((?x, \textit{p}, ?y)\ \OPT\ (?x, \textit{q}, ?z))\ANDAND (?x,
\textit{r}, ?z))$ is not in OPT normal form.

Note that all patterns in OPT normal form have the following form:
\begin{equation}
P_0 \ \OPT\ P_1\ \OPT \ldots \OPT \ P_m \footnote{We abbreviate $((P_0 \ \OPT\ P_1)\ \OPT \ldots \OPT \ P_m)$ as $P_0 \ \OPT\ P_1\ \OPT \ldots \OPT \ P_m$.};
\end{equation}
where $P_0$ is an OPT-free pattern, that is, $P_0$ contains only
AND and FILTER operations (called $AF$-pattern). In this sense, we
use $BGP(P)$ to denote $P_0$ and $\mathcal{O}(P)$ to denote $\{P_1, \ldots,
P_m\}$, i.e., the collection of optional patterns occurring in $P$.

\begin{prop}\cite[Theorem 4.11]{P2009Semantics}\label{prop:ONF}
For every UNION-free well-designed pattern $P$, there exists a pattern
$Q$ in OPT normal form such that $P$ and $Q$ are equivalent.
\end{prop}

In the proof of Proposition \ref{prop:ONF}, we apply three rewriting rules
based on the following equations: let $P, Q, R$ be patterns and $C$ a
constraint,
\begin{compactitem}
\item $(P \OPT R) \FILTER C \equiv (P \FILTER C) \OPT R$;
\item $(P \OPT R) \ANDAND Q \equiv (P \ANDAND Q) \OPT R$;
\item $P \ANDAND (Q \OPT R) \equiv (P \ANDAND Q) \OPT R$.
\end{compactitem}

Since each UNION-free well-designed pattern is equivalent to a pattern in
OPT normal form by Proposition \ref{prop:ONF}, we mainly consider all
well-designed patterns in OPT normal form in the following.

\begin{figure}[H]
\begin{center}
\begin{tikzpicture}
  \matrix (m) [matrix of math nodes,row sep=3em,column sep=4em,minimum width=2em]
  {
     t_1 & t_2 & t_4 \\
      & t_3 & t_5\\};
  \path[-stealth]
    (m-1-1) edge node [above] {$\OPT$} (m-1-2)
    (m-1-2) edge node [above] {$\OPT$} (m-1-3)
    (m-1-3) edge node [right] {$\OPT$} (m-2-3)
    (m-1-2) edge node [right] {$\OPT$} (m-2-2);
\end{tikzpicture}
\caption{The figure of OPT normal form}\label{Fig:opt-normal-form}
\end{center}
\end{figure}
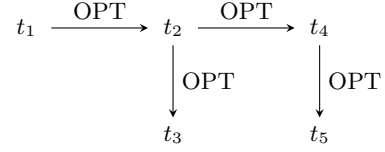

To further observe some features of patterns in OPT normal form, we consider
a complicated pattern $P$, where the OPT operation is deeply nested, as
follows:
\begin{equation}\label{equ:ONF}
P= (t_1\ \OPT \ (t_2 \ \OPT \ t_3))\ \OPT \ (t_4 \ \OPT \ t_5).
\end{equation}
Note that, in $P$, $t_1$ is non-optional while $t_2, t_3, t_4$ and $t_5$
are optional. Furthermore, if we consider the subpattern $(t_2 \ \OPT \
t_3)$, $t_2$ is non-optional while $t_3$ is still optional. Analogously, if
we consider the subpattern $(t_4 \ \OPT \ t_5)$, $t_4$ is non-optional while
$t_5$ is still optional. Now, if we observe the
figure of $P$ shown in Figure \ref{Fig:opt-normal-form}, $t_2$ and $t_4$ are
on top of $t_3$ and $t_4$, respectively.

\vspace{-4mm}

\subsection{OPT-depth in OPT Normal Form}
To characterize the different levels of optional patterns, we define
\emph{OPT-depth} of patterns in OPT normal form.

\begin{defn}[OPT-depth]\label{def:OPT-depth}
Let $P$ be a pattern in OPT normal form. We use $\depth{P}$ to denote its
\emph{OPT-depth} as follows:
\begin{compactitem}
\item $\depth{P} = 0$ if $P$ is an $AF$-pattern;
\item  $\depth{P} = \max\{\depth{P_1},\ldots, \depth{P_m}\}+1$ if
    $\mathcal{O}(P) = \{P_1, \ldots, \ P_m\}$.
\end{compactitem}
\end{defn}

For instance, the OPT-depth of the pattern $Q$ stated in Section \ref{sec:introduction} and
the pattern $P$ in Equation (\ref{equ:ONF}) are 2.

\subsection{Approximate Queries}
To define our approximate queries, we introduce an important notion called
\emph{reduction} \cite{P2009Semantics}.

We say that a pattern $P'$ is a \emph{reduction} of a pattern $P$, if $P'$ can
be obtained from $P$ by replacing subpattern $(P_1\ \OPT\ P_2)$ with $P_1$,
that is, $P'$ is obtained by deleting some optional parts of $P$.
The reflexive and transitive closure of the reduction relation is denoted by
$\unlhd$. In this sense, for a pattern, its reductions can be taken as ``inexact'' patterns,
which can be obtained by reducing the OPT operation.
For instance, in Section \ref{sec:introduction}, $Q_1$ and $Q_2$ are reductions of $Q$.

Inspired from the notion of reduction, we introduce our \emph{k-approximate
patterns}.
\begin{defn}[k-approximation]
Let $P$ be a pattern in OPT normal form ($P_0 \ \OPT \ P_1 \ \OPT \ldots
\OPT \ P_m$) and $k$ be a natural number. The \emph{k-approximate pattern} of
$P$ (written as $P^{(k)}$) can be obtained in the following inductive way:
\begin{compactitem}
\item $P^{(k)} = \BGP{P}$ if $k=0$;
\item $P^{(k)}= P_0 \OPT P_1^{(k-1)} \ \OPT \ldots \OPT P_m^{(k-1)}$ if $1
    \le k \le \depth{P}-1$;
\item $P^{(k)}= P$ if $k \ge \depth{P}$.
\end{compactitem}
\end{defn}

Intuitively, approximate patterns are subpatterns obtained by reducing their
OPT-depths. In this sense, our approximation generalizes reduction
\cite{P2009Semantics} in a fine-grained way.
Since there exists the unique OPT-depth for each OPT in OPT normal form, we have the following proposition:

\begin{prop}\label{prop:app-unqiue-2}
Let $P$ be a pattern in OPT normal form and $k$ be a natural number. $P^{(k)}$
exists and $P^{(k)}$ is unique.
\end{prop}

For instance, in Section \ref{sec:introduction}, $Q^{(0)} = Q_1$ and $Q^{(1)} = Q_2$. In Equation
(\ref{equ:ONF}), $P^{(0)} = t_1$ and $P^{(1)} = ((t_1 \ \OPT \ t_2)\ \OPT \ t_4)$.
$Q^{(0)}$ and $Q^{(1)}$ are the reductions of $Q$. Analogously, $P^{(0)}$ and
$P^{(1)}$ are the reductions of $P$.

\section{K-approximation computation}\label{sec:well-designed tree}

In this section, we propose a method to compute all approximate
patterns based on a redesigned parse tree called \emph{well-designed tree}.

Now, we introduce the notion of \emph{well-designed tree}.
\begin{defn}[well-designed tree]\label{well-designed tree}
Let $P$ be a well-designed pattern in OPT normal form. A well-designed tree $T$ based on $P$ is a redesigned parse tree,
which can be defined as follows:
\begin{compactitem}
\item All inner nodes in $T$ are labeled by $\OPT$ operations and leaf nodes are
    labeled by $AF$-patterns.
\item For each subpattern $(P_1\ \OPT \ P_2)$ of $P$, the well-designed
    tree $T_1$ of $P_1$ and the well-designed tree $T_2$ of $P_2$ have the same parent node.
\end{compactitem}
\end{defn}

\begin{figure}[ht]\centering
\caption{Well-designed Tree\label{fig:wd}}
\begin{tikzpicture}[scale=0.5][nodes={draw}, -]
\node{$\OPT_1$}
  child { node {$\OPT_2$}
    child { node {$p_0$}}
    child { node {$t_2$} }
  }
  child [missing]
  child [missing]
  child { node {$\OPT_3$}
    child { node {$\OPT_4$}
        child { node {$t_4$}}
        child { node {$t_5$}}
    }
  child [missing]
    child { node {$\OPT_5$}
        child { node {$t_6$}}
        child { node {$t_7$}}
    }
  };
\end{tikzpicture}
\end{figure}
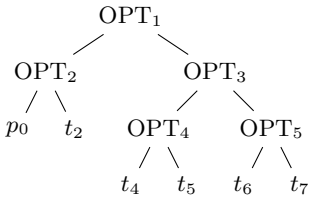

For instance, given a pattern $P$\footnote{We give each OPT operator a
subscript to differentiate them so that readers understand clearly.} in OPT
normal form,
\begin{multline*}
P=((((t_1 \ \ANDAND \ t_3) \ \FILTER \ C) \ \OPT_2 \  t_2) \ \OPT_1 \ \ \ \ \  \ \ \  \ \ \  \ \ \  \\ ((t_4 \ \OPT_4 \ t_5) \ \OPT_5 (t_6 \ \OPT_6 \ t_7))).
\end{multline*}
We write $((t_1 \ \ANDAND \ t_3) \ \FILTER \ C)$ as $p_0$ for short, which is the non-optional part of $P$.
The well-designed tree $T$ is shown in Figure \ref{fig:wd}.
\vspace{-4mm}
\vspace{-3mm}

Some pruning strategies can be applied to the well-designed tree to achieve
$k$-approximation. After removing optional subtrees from the well-designed tree,
we get a $k$-approximation spanning tree ($\kst$ for short) which is also a
well-designed tree.
We denote a $k$-approximation spanning tree from well-designed
tree $T$ as $\kst_T^{(k)}$.
In order to obtain $\kst_T^{(k)}$, we define a
special traversal method for the well-designed tree based on the conception of OPT-depth, called \emph{Left-Deep Level Traversal}. Before defining \emph{Left-Deep Level Traversal},
we provide a partial traversal approach called \emph{Leftmost Traversal}.

For a well-designed tree, \emph{Leftmost Traversal} of this tree is by only traversing the left subtree after visiting root node.
For instance, consider $T$ in Figure \ref{fig:wd}, the leftmost traversal of $T$ is denoted by
$\LT{T}=\{\OPT_1, \OPT_2, p_0\}$.
\emph{Left-Deep Level Traversal} of the well-designed tree is proposed as follows:
\begin{defn}[left-deep level traversal]
Let $T$ be a well-designed tree.
Left-Deep Level Traversal denoted by $\ld(T)$ is composed of levels.
$\lev{i}$ can be obtained by leftmost traversing each node's right children node (called candidate) in $\lev{i-1}$.
Especially, $\lev{0} = \LT{T}$.
\end{defn}

For each subtree $t$ in the well-designed tree, the leftmost leaf node written as
$\LM{t}$ is the non-optional part of $t$. For instance, for the
well-designed tree $T$ in Figure \ref{fig:wd}, $\LM{T}=\{p_0\}$.
We construct $\kst_T^{(k)}$ by removing the subtrees below $\lev{k-1}$ from $T$.
Particularly, $\kst_T^{(0)}$ can be built by returning $\LM{T}$.

In the process of building $\kst_T^{(k)}$, firstly we compute each node's candidate in $\lev{k-1}$. Secondly we obtain the $\LM{n}$ for each OPT node $n$ in $\lev{k-1}$. Finally $\kst_T^{(k)}$ can be constructed by replacing the leftmost nodes with corresponding OPT nodes in $T$.
We obtain the $k$-approximation query through traversing on $\kst_T^{(k)}$.
The process of building $\kst_T^{(k)}$ is described in Algorithm \autoref{kst}.

\begin{exmp}
Consider the well-designed tree $T$ in Figure \ref{fig:wd} from pattern $P$.
The $\ld(T)$\footnote{We use $\times$ to denote that for each non-OPT node $n$ in candidates, there exist no corresponding $\LM{n}$ in leftmost list.} with candidates and leftmost list can be described as follows:
\begin{center}
\begin{tabular}{|c|l|l|l|}
\hline
Level &Traversal List &Candidates &Leftmost \\ \hline
0 &$\OPT_1$, $\OPT_2$, $p_0$ &$\OPT_3$, $t_2$ &$t_4$, $\times$ \\ \hline
1 &$\OPT_3$, $\OPT_4$, $t_4$, $t_2$ &$\OPT_5$, $t_5$ &$t_6$, $\times$ \\ \hline
2 &$\OPT_5$, $t_6$, $t_5$ &$t_7$ &$\times$ \\ \hline
3 &$t_7$ &  & \\ \hline
\end{tabular}
\end{center}
In $\kst_T^{(0)}$, $p_0$ is set as the root node without any child node.
If we want to obtain $\kst_T^{(1)}$, we can replace $t_4$ with $\OPT_3$ in $T$ based on $\lev{0}$.
Analogously, $\kst_T^{(2)}$ can be obtained by replacing $t_6$ with $\OPT_5$ in $T$ based on $\lev{1}$.
Since $\depth{P} = 3$, $\kst_T^{(3)}$ is regarded as $T$ itself.
Both $\kst_T^{(1)}$ and $\kst_T^{(2)}$ are shown in Figure \ref{fig:reduction}.
\vspace{-4mm}
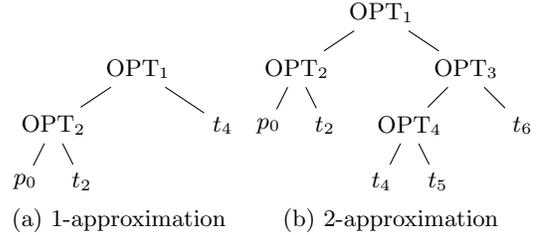
\begin{figure}[h]\centering
\caption{Approximation Spanning Tree\label{fig:reduction}}
\subfigure[1-approximation]{ \label{fig:1-reduction}
\begin{tikzpicture}[scale=0.5][nodes={draw}, -]
\node{$\OPT_1$}
  child { node {$\OPT_2$}
    child { node {$p_0$}}
    child { node {$t_2$} }
  }
    child [missing]
  child [missing]
  child { node {$t_4$}};
\end{tikzpicture}}\subfigure[2-approximation]{ \label{fig:2-reduction}
\begin{tikzpicture}[scale=0.5][nodes={draw}, -]
\node{$\OPT_1$}
  child { node {$\OPT_2$}
    child { node {$p_0$}}
    child { node {$t_2$} }
  }
  child [missing]
  child [missing]
  child { node {$\OPT_3$}
    child { node {$\OPT_4$}
        child { node {$t_4$}}
        child { node {$t_5$}}
    }
  child [missing]
    child { node {$t_6$}}
  };
\end{tikzpicture}}
\end{figure}
\\
$[P]^1$ and $[P]^2$ are shown as follows:
\begin{multline*}
[P]^1=((((t_1 \ \ANDAND \ t_3) \ \FILTER \ C) \ \OPT_2 \ t_2) \ \OPT_1 \ t_4),
\end{multline*}
and
\begin{multline*}
[P]^2=((((t_1 \ \ANDAND \ t_3) \ \FILTER \ C) \ \OPT_2 \ t_2) \ \OPT_1 \ \ \ \ \  \ \ \  \\ \ \ \   \ \ \ \ \ \ \ \ \ \ \ \ \ \ \ \ \ \ \ \ \ \ \ \ \ \ \ \ \ \ \ \ \ \ \ \ \ \ \ \ \ \ \ \ \ \ \ \ \ \ \ \ \ \ ((t_4 \ \OPT_4 \ t_5) \ \OPT_5 \ t_6)).
\end{multline*}
\end{exmp}

\begin{center}
\begin{algorithm}[ht]
\caption{$K$-approximation Spanning Tree \label{kst}}
\begin{algorithmic}[1]
\REQUIRE    Well-designed tree $T$ from pattern $P$, Leftmost list $leftmost$,
    and $k$-approximation with $k$\\
    Initialize Candidate $candidate$ with $T$, $i \leftarrow 0$
\ENSURE     $K$-approximation Spanning Tree
\IF{$k = 0$}
\RETURN $\LM{T}$
\ELSIF{$k \geq  \depth{P}$}
\RETURN $T$
\ELSE
\WHILE{$i \neq k$}
    \STATE $\lev{i} \leftarrow \LT{candidate}$
    \STATE $candidate \leftarrow GetCandidate(\lev{i})$.
    \FOR{each $node$ in $candidate$}
        \IF{$node$ is $\OPT$}
        \STATE $leftmost$ $\leftarrow$ $\LM{node}$
        \ENDIF
    \ENDFOR
\ENDWHILE
\STATE Replace the nodes in $leftmost$ with corresponding OPT nodes in $T$.
\RETURN $T$
\ENDIF
\end{algorithmic}
\end{algorithm}
\end{center}

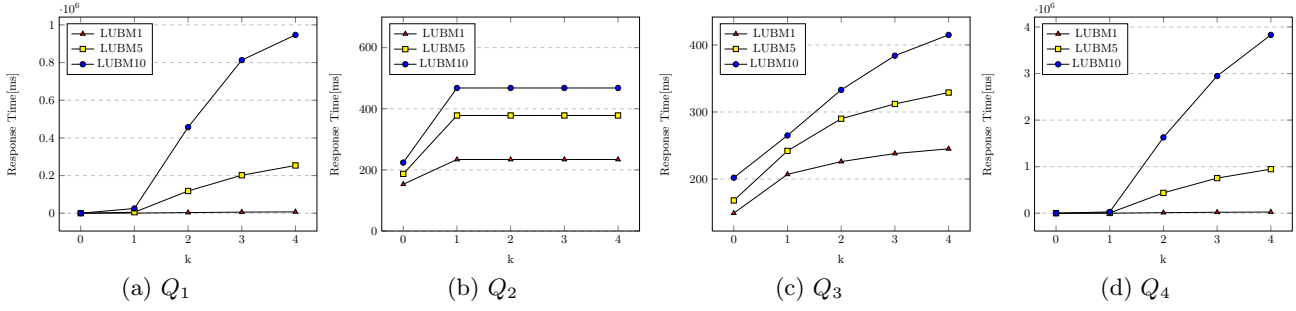
\begin{figure*}[ht]\centering
\caption{K-approximation on Jena \label{jena}}
\subfigure[$Q_1$] {
\begin{tikzpicture}[scale=0.5]
\begin{axis}[
    xlabel={k},
    ylabel={Response Time[ms]},
    symbolic x coords={0,1,2,3,4},
    legend pos=north west,
    ymajorgrids=true,
    grid style=dashed,
]

\addplot[
    color=black,
    mark=triangle*, mark options={fill=red}
    ]
    coordinates {
    (0,144) (1,747) (2,3404) (3,5968) (4,7207)
    };
\addplot[
    color=black,
    mark=square*, mark options={fill=yellow}
    ]
    coordinates {
    (0,162) (1,5473) (2,118129) (3,201746) (4,253567)
    };
\addplot[
    color=black,
    mark=*, mark options={fill=blue}
    ]
    coordinates {
    (0,212) (1,25190) (2,456717) (3,812985) (4,946881)
    };
    \legend{LUBM1,LUBM5,LUBM10}

\end{axis}
\end{tikzpicture}}\subfigure[$Q_2$] {
\begin{tikzpicture}[scale=0.5]
\begin{axis}[
    xlabel={k},
    xtick=data,
    ylabel={Response Time[ms]},
    ymin=0,
    ymax=700,
    symbolic x coords={0,1,2,3,4},
    legend pos=north west,
    ymajorgrids=true,
    grid style=dashed,
]

\addplot[
    color=black,
    mark=triangle*, mark options={fill=red}
    ]
    coordinates {
    (0,153) (1,234)(2,234)(3,234)(4,234)
    };
\addplot[
    color=black,
    mark=square*, mark options={fill=yellow}
    ]
    coordinates {
    (0,187) (1,378) (2,378) (3,378) (4,378)
    };
\addplot[
    color=black,
    mark=*, mark options={fill=blue}
    ]
    coordinates {
    (0,224) (1,468) (2,468) (3,468) (4,468)
    };
    \legend{LUBM1,LUBM5,LUBM10}

\end{axis}
\end{tikzpicture}
}\subfigure[$Q_3$] {
\begin{tikzpicture}[scale=0.5]
\begin{axis}[
    xlabel={k},
    ylabel={Response Time[ms]},
    symbolic x coords={0,1,2,3,4},
    legend pos=north west,
    ymajorgrids=true,
    grid style=dashed,
]

\addplot[
    color=black,
    mark=triangle*, mark options={fill=red}
    ]
    coordinates {
    (0,149) (1,207) (2,226) (3,238) (4,245)
    };
\addplot[
    color=black,
    mark=square*, mark options={fill=yellow}
    ]
    coordinates {
    (0,168) (1,242) (2,290) (3,312) (4,329)
    };
\addplot[
    color=black,
    mark=*, mark options={fill=blue}
    ]
    coordinates {
    (0,202) (1,265) (2,333) (3,384) (4,415)
    };
    \legend{LUBM1,LUBM5,LUBM10}

\end{axis}
\end{tikzpicture}}\subfigure[$Q_4$] {
\begin{tikzpicture}[scale=0.5]
\begin{axis}[
    xlabel={k},
    xtick=data,
    ylabel={Response Time[ms]},
    symbolic x coords={0,1,2,3,4},
    legend pos=north west,
    ymajorgrids=true,
    grid style=dashed,
]

\addplot[
    color=black,
    mark=triangle*, mark options={fill=red}
    ]
    coordinates {
    (0,156) (1,797) (2,11935) (3,20458) (4,25733)
    };
\addplot[
    color=black,
    mark=square*, mark options={fill=yellow}
    ]
    coordinates {
    (0,175) (1,5698) (2,436956) (3,754456) (4,946314)
    };
\addplot[
    color=black,
    mark=*, mark options={fill=blue}
    ]
    coordinates {
    (0,220) (1,27802) (2,1628827) (3,2946824) (4,3830155)
    };
    \legend{LUBM1,LUBM5,LUBM10}

\end{axis}
\end{tikzpicture}
}
\end{figure*}
\begin{figure*}[ht]\centering
\caption{K-approximation on Sesame \label{sesame}}
\subfigure[$Q_1$] {
\begin{tikzpicture}[scale=0.5]
\begin{axis}[
    xlabel={k},
    ylabel={Response Time[ms]},
    symbolic x coords={0,1,2,3,4},
    legend pos=north west,
    ymajorgrids=true,
    grid style=dashed,
]

\addplot[
    color=black,
    mark=triangle*, mark options={fill=red}
    ]
    coordinates {
    (0,73) (1,372) (2,2750) (3,4269) (4,6357)
    };
\addplot[
    color=blue,
    mark=square*, mark options={fill=yellow}
    ]
    coordinates {
    (0,125) (1,4128) (2,78711) (3,196800) (4,246231)
    };
\addplot[
    color=blue,
    mark=*, mark options={fill=blue}
    ]
    coordinates {
    (0,145) (1,20061) (2,339523) (3,765155) (4,843060)
    };
    \legend{LUBM1,LUBM5,LUBM10}

\end{axis}
\end{tikzpicture}}\subfigure[$Q_2$] {
\begin{tikzpicture}[scale=0.5]
\begin{axis}[
    xlabel={k},
    xtick=data,
    ylabel={Response Time[ms]},
    ymin=0,
    ymax=700,
    symbolic x coords={0,1,2,3,4},
    legend pos=north west,
    ymajorgrids=true,
    grid style=dashed,
]

\addplot[
    color=black,
    mark=triangle*, mark options={fill=red}
    ]
    coordinates {
    (0,81) (1,140)(2,140)(3,140)(4,140)
    };
\addplot[
    color=black,
    mark=square*, mark options={fill=yellow}
    ]
    coordinates {
    (0,103) (1,251) (2,251) (3,251) (4,251)
    };
\addplot[
    color=black,
    mark=*, mark options={fill=blue}
    ]
    coordinates {
    (0,147) (1,372) (2,372) (3,372) (4,372)
    };
    \legend{LUBM1,LUBM5,LUBM10}

\end{axis}
\end{tikzpicture}
}\subfigure[$Q_3$] {
\begin{tikzpicture}[scale=0.5]
\begin{axis}[
    xlabel={k},
    ylabel={Response Time[ms]},
    symbolic x coords={0,1,2,3,4},
    legend pos=north west,
    ymajorgrids=true,
    grid style=dashed,
]

\addplot[
    color=black,
    mark=triangle*, mark options={fill=red}
    ]
    coordinates {
    (0,74) (1,102) (2,123) (3,139) (4,154)
    };
\addplot[
    color=black,
    mark=square*, mark options={fill=yellow}
    ]
    coordinates {
    (0,119) (1,162) (2,198) (3,229) (4,256)
    };
\addplot[
    color=black,
    mark=*, mark options={fill=blue}
    ]
    coordinates {
    (0,136) (1,252) (2,310) (3,356) (4,364)
    };
    \legend{LUBM1,LUBM5,LUBM10}

\end{axis}
\end{tikzpicture}}\subfigure[$Q_4$\label{fig:q4-2}] {
\begin{tikzpicture}[scale=0.5]
\begin{axis}[
    xlabel={k},
    xtick=data,
    ylabel={Response Time[ms]},
    symbolic x coords={0,1,2,3,4},
    legend pos=north west,
    ymajorgrids=true,
    grid style=dashed,
]

\addplot[
    color=black,
    mark=triangle*, mark options={fill=red}
    ]
    coordinates {
    (0,80) (1,513) (2,8130) (3,16742) (4,22872)
    };
\addplot[
    color=black,
    mark=square*, mark options={fill=yellow}
    ]
    coordinates {
    (0,114) (1,4523) (2,274395) (3,704830) (4,911398)
    };
\addplot[
    color=black,
    mark=square*, mark options={fill=blue}
    ]
    coordinates {
    (0,131) (1,25204) (2,1127656) (3,2842586) (4,3673732)
    };
    \legend{LUBM1,LUBM5,LUBM10}

\end{axis}
\end{tikzpicture}
}
\end{figure*}
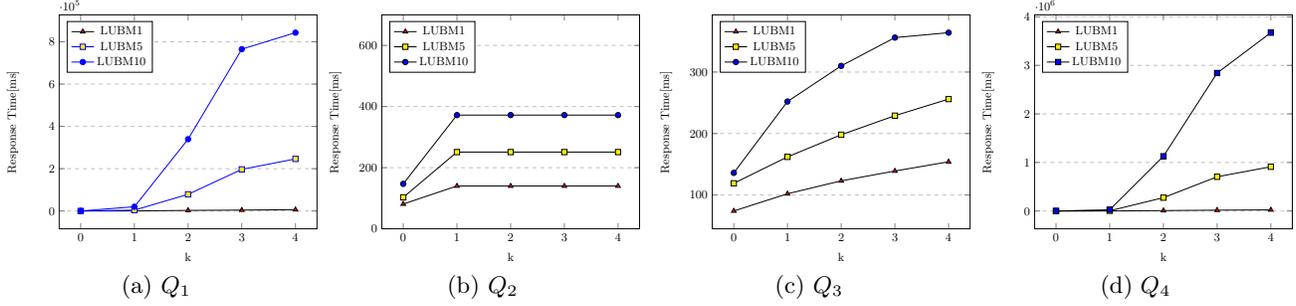

\section{Experiments and evaluations}\label{sec:implement}

This section presents our experiments.
The purpose of the experiments is to evaluate (1) the performance improvement
of approximate well-designed SPARQL queries, and (2) the appropriate $k$ to reduce the users' waiting time for solutions.

\subsection{Experiments}
\paragraph{Implementations and running environment} All experiments were carried
out on a machine running Linux, which has one CPU with four cores of
2.40GHz, 32GB memory and 500GB disk storage. All of the algorithms were
implemented in JAVA with Eclipse as our compiler. Jena\cite{Carroll2004Jena}
(Jena-3.0.1) and Sesame\cite{Broekstra2002Sesame} (Sesame-4.1.1) are used as
the underlying query engines of approximate queries.

\paragraph{Dataset}
We used LUBM\footnote{http://swat.cse.lehigh.edu/projects/lubm/} as the
dataset in our experiments to look for the relationship between approximate query
efficiency and $k$. LUBM, which features an ontology for the university domain, is a standard benchmark
to evaluate the performance of Semantic Web repositories,
In our experiments, we used LUBM1, LUBM5 and
LUBM10 as query datasets shown in Table \ref{tab:lubm}.
\vspace{-3mm}
\begin{table}[h]
\centering
\caption{Profiles of datasets \label{tab:lubm}}
\begin{tabular}{|c|r|r|}  
\hline
Dataset  &Number of Triples  &NT File Size(bytes)\\ \hline  
LUBM1 &103,104 &14,497,954\\ \hline         
LUBM5 &645,836 &90,960,405\\  \hline
LUBM10 &1,316,701 &185,474,846\\       
\hline
\end{tabular}
\end{table}
\vspace{-3mm}
\paragraph{SPARQL queries} The queries over LUBM were designed as 4 forms
in Table \ref{tab:query}. Obviously, OPT nesting in $Q_4$ is the most complex among 4 forms.
Furthermore, we built $\ANDAND$ and $\FILTER$ operations in each query.
All of query patterns have $k$ ranging from 0 to 4. Specially, since
$\depth{Q_2}$ is 1, we regard k-approximate query as $Q_2$ itself when
$k>1$.
\vspace{-3mm}
\begin{table}[h]
\centering
\caption{Queries on LUBM \label{tab:query}}
\begin{tabular}{|c|l|c|}  
\hline
Query  &Well-designed tree  &Amount of OPT\\ \hline
$Q_1$ &zigzag tree &9\\ \hline
$Q_2$ &left-deep tree &4\\  \hline
$Q_3$ &right-deep tree &4\\ \hline
$Q_4$ &full tree &15\\ \hline
\end{tabular}
\end{table}
\vspace{-3mm}
\paragraph{The amount of OPT after approximation}
The amount of OPT with different $k$ is shown in Table \ref{tab:amount}.
Clearly, the amount of OPT is decreasing after approximation since our approximation method
can reduce OPT-depth.
Note that when k is 4, query is itself without any approximation.
\vspace{-3mm}
\begin{table}[h]
\centering
\caption{Amount of OPT after approximation \label{tab:amount}}
\begin{tabular}{|c|c|c|c|c|c|}  
\hline
k     &k=0  &k=1 &k=2 &k=3 &k=4\\ \hline
$Q_1$ &0 &2 &5 &8 &9\\ \hline
$Q_2$ &0 &4 &4 &4 &4\\  \hline
$Q_3$ &0 &1 &2 &3 &4\\ \hline
$Q_4$ &0 &4 &10 &14 &15\\ \hline
\end{tabular}
\end{table}
\vspace{-3mm}

\subsection{Efficiency of Approximate Queries}
For a well-designed query $Q$ and its $k$-approximation
query $Q^{(k)}$, $Q^{(k)}$ is more closed to $Q$ with higher value of $k$.
The variation tendencies of query response time shown in Figure \ref{jena} and
Figure \ref{sesame} are similar.
Query efficiency is promoted with lower
response time when $k$ is decreasing (approximation degree becomes larger).
Furthermore, there has been a significant increase in query efficiency when the dataset scale grows up.
For instance, we observe $Q_4$, which corresponds to a full well-designed
tree. When the dataset is LUBM10, its query response time is more than an
hour implemented by Jena and Sesame without any approximation ($Q_4^{(4)}$). However, the
response time of $Q_4^{(1)}$ is less than a minute. Furthermore, comparing
$Q_4^{(3)}$ with $Q_4^{(4)}$ implemented by Jena and Sesame, an approximately decrease of 25\% in
the query response time has shown in both Figure \ref{jena} and Figure \ref{sesame}. $Q_2$ and $Q_3$
has less time than $Q_1$ and $Q_4$ since $Q_2$ and $Q_3$ have less OPT amounts and simpler OPT nestings.

Approximate queries can efficiently reduce the query response time and users' waiting time.
An appropriate $k$ can be determined to reduce users' waiting time for solutions since users' tolerable waiting time is limited.
We assume that $Q_4$ on LUBM10 comes from users,
and it takes more than an hour time to answer $Q_4^{(4)}$ by Jena and Sesame if users want to obtain all exact solutions,
which might lead to bad user experience.
In this scene, it can be approximated as $Q_4^{(1)}$ to improve user experience within a minute waiting time.

More results of k-approximation can be found in the online demo website:
\url{http://123.56.79.184/approximate.html}.

\section{Conclusion}\label{sec:con}
In this paper, we have presented the approximation of well-designed SPARQL
patterns in OPT normal form based on the depth of OPT operation.
Theoretically, our proposal k-approximation generalizes reductions of
patterns in a fine-grained way. The k-approximation provides rich and
various approximate queries to answer user's query within a fixed time.
Our experimental results show that our approximation on the depth of OPT operation is reasonable and useful.

In the future, we are going to handle other non-well-designed patterns and
deal with more operations such as UNION. Besides, we will extend the
approximation method to obtain other approximation queries.

\section{Acknowledgments}
This work is supported by the program of the National Natural Science
Foundation of China (NSFC) under 61502336, 61572353, 61373035 and the
National High-tech R\&D Program of China (863 Program) under 2013AA013204.
Xiaowang Zhang is supported by the project-sponsored by School of Computer
Science and Technology in Tianjin University.

\balance

\begin{thebibliography}{10}

\bibitem{gutierrez_survey} R.~Angles and C.~Gutierrez.
\newblock Survey of graph database models.
\newblock {\em ACM Computing Surveys}, 40(1)(2008): article 1.

\bibitem{Abiteboul1995Foundations} S.~Abiteboul, H.~Richard, and
    V.~Vianu.
\newblock Foundations of databases.
\newblock {\em Addison Wesley}, page 9:3¨C9:56, 1995.

\bibitem{abiteboul2000data} S.~Abiteboul, P.~Buneman, and D.~Suciu.
\newblock {\em Data on the Web: from relations to semistructured data and XML}.
\newblock Morgan Kaufmann, 2000.

\bibitem{barcelo2015efficient} P.~Barcelo, R.~Pichler, and S.~Skritek.
\newblock Efficient evaluation and approximation of well-designed pattern
  trees.
\newblock In {\em Proc. of PODS 2015}, pages 131--144. ACM, 2015.

\bibitem{Broekstra2002Sesame} J.~Broekstra, A.~Kampman, and F.~V. Harmelen.
\newblock {\em Sesame: A generic architecture for storing and querying RDF and
  RDF Schema}.
\newblock Springer Berlin Heidelberg, 2002.

\bibitem{Carroll2004Jena} J.~J. Carroll, I.~Dickinson, C.~Dollin,
    D.~Reynolds, A.~Seaborne, and K.~Wilkinson.
\newblock Jena: implementing the semantic web recommendations.
\newblock In {\em Proc. of WWW 2004}, pages 74--83, 2004.


\bibitem{Klyne2004Resource} G.~Klyne, C.~J.~Jeremy, and B.~McBride.
\newblock Resource description framework (RDF): Concepts and abstract syntax.
\newblock {\em W3C Recommendation}, 2004.

\bibitem{rafragments} G.H.L. Fletcher, M.~Gyssens,
    D.~Leinders, J.~Van~den Bussche, D.~Van~Gucht,
  S.~Vansummeren, and Y.~Wu.
\newblock Relative expressive power of navigational querying on graphs.
\newblock In {\em Proc. of ICDT 2011}, pp.197--207, 2011.

\bibitem{Letelier2012Static} A.~Letelier, J.~P¨¦rez, R.~Pichler, and
    S.~Skritek.
\newblock Static analysis and optimization of semantic web queries.
\newblock In {\em Proc. of PODS 2012}, 38(4):84--87, 2012.

\bibitem{Nah2003A} F.~H. Nah.
\newblock A study on tolerable waiting time: How long are web users willing to
  wait?
\newblock {\em Behaviour and Information Technology}, 23(3), 2003.

\bibitem{walklogic} J.~Hellings, B.~Kuijpers, J.~Van~den Bussche, and
    X.~Zhang.
\newblock Walk logic as a framework for path query languages on graph
  databases.
\newblock In {\em  Proc. of ICDT 2013}, pp.117--128, 2011.

\bibitem{P2009Semantics} J.~P¨¦rez, M.~Arenas, and C.~Gutierrez.
\newblock Semantics and complexity of SPARQL.
\newblock {\em ACM Transactions on Database Systems}, 34(3):30--43, 2009.

\bibitem{Prud2007SPARQL} E.~Prud'Hommeaux and A.~Seaborne.
\newblock SPARQL query language for RDF.
\newblock {\em W3C Recommendation}, 2008.

\bibitem{saleem2015lsq} M.~Saleem, M.~I. Ali, A.~Hogan, Q.~Mehmood, and
    A.-C.~N. Ngomo.
\newblock LSQ: The linked SPARQL queries dataset.
\newblock In {\em Proc. of ISWC 2015}, pages 261--269. Springer, 2015.

\bibitem{schmidt_sparqloptim} M.~Schmidt, M.~Meier, and G.~Lausen.
\newblock Foundations of {SPARQL} query optimization.
\newblock In {\em Proc. of ICDT'10}, pp. 4--33, 2010.

\bibitem{wood_survey} P.~Wood,
\newblock Query languages for graph databases,
\newblock {\em SIGMOD Record}, 41(1) (2012): 50--60.

\bibitem{Zhang2015sat} X.~Zhang and J.~Van~den. Bussche.
\newblock On the satisfiability problem for SPARQL patterns.
\newblock {\em arXiv:1406.1404}, 2014.

\bibitem{Zhang2014On} X.~Zhang and J.~Van~den. Bussche.
\newblock On the primitivity of operators in SPARQL.
\newblock {\em Information Processing Letters}, 114(9):480--485, 2014.

\end{thebibliography}

\end{document}